\documentclass[aps,prb,twocolumn,superscriptaddress,showpacs]{revtex4}

\usepackage{bm}
\usepackage[dvips]{graphicx}

\begin{document}

\title{Spontaneous rearrangement of the checkerboard charge order \\
to stripe order in La$_{1.5}$Sr$_{0.5}$NiO$_{4}$}

\date{\today}

\author{R. Kajimoto}
\affiliation{Department of Physics, Ochanomizu University, Bunkyo-ku,
Tokyo 112-8610, Japan}

\author{K. Ishizaka}
\affiliation{Department of Applied Physics, University of Tokyo,
Bunkyo-ku, Tokyo 113-8656, Japan}

\author{H. Yoshizawa}
\affiliation{Neutron Scattering Laboratory, ISSP, University of Tokyo,
Tokai, Ibaraki 319-1106, Japan}

\author{Y. Tokura}
\affiliation{Department of Applied Physics, University of Tokyo,
Bunkyo-ku, Tokyo 113-8656, Japan}

\begin{abstract}

 The charge ordering in La$_{1.5}$Sr$_{0.5}$NiO$_{4}$ has been studied
 with neutron diffraction technique. An interesting rearrangement of the
 charge ordering was observed as a function of temperature. With
 decreasing temperature, a checkerboard-type charge order is formed
 below $T_\mathrm{CO}^\mathrm{C} \sim 480$ K. It is, however, taken over
 by the stripe-type charge order with an incommensurability twice as
 large as that of the spin order below $T_\mathrm{CO}^\mathrm{IC} \sim
 180$ K.  Surprisingly, the stripe phase persists up to $x = 0.7$ for
 highly hole-doped samples of Nd$_{2-x}$Sr$_{x}$NiO$_{4}$ with $0.45
 \leq x \leq 0.7$.

\end{abstract}


\maketitle

%\section{Introduction}

One of the most fascinating examples of charge ordering may be a
so-called stripe order discovered in the high-$T_c$ superconducting
cuprates.\cite{tranquada95} Recent studies have revealed that the stripe
order cannot be simply attributed to the Coulomb interactions, but is a
consequence of an interplay among spin, charge, and lattice degrees of
freedom. A stripe order is also known to exist in the isostructural
nickelates, although its stripe order is relatively static compared with
that of the cuprates.  In hole-doped nickelates
La$_{2-x}$Sr$_{x}$NiO$_{4+\delta}$, the static stripe order is observed
for a wide hole concentration region $0.15 \lesssim n_\mathrm{h} \le
1/2$,\cite{tranquada96,yoshizawa00} where $n_\mathrm{h} = x + 2\delta$.
In the stripe ordered state, the doped holes form charge stripes
characterized by a modulation vector $\bm{g}_\mathrm{CO} =
(2\epsilon,0,0)$ in the orthorhombic lattice units ($a_\mathrm{o} =
\sqrt{2}a_\mathrm{t}$). The Ni spins order antiferromagnetically with
antiphase domain boundaries on the charge stripes. The modulation vector
of the spin order thus becomes $\bm{g}_\mathrm{SO} =
\bm{Q}_\mathrm{AF}\pm (\epsilon,0,0)$, where $\bm{Q}_\mathrm{AF} =
(1,0,0)$ is the wave vector for a simple antiferromagnetic order. The
incommensurability $\epsilon$ keeps a linear relation to $n_\mathrm{h}$
for $n_\mathrm{h} \le 1/2$, $\epsilon \sim
n_\mathrm{h}$.\cite{tranquada96,yoshizawa00} The stripe order is most
stable at $\epsilon=1/3$ where it shows the highest charge and spin
ordering temperatures and the longest correlation length. Moreover,
$\epsilon$ exhibits a systematic deviation from the linear relation
around $n_\mathrm{h} \sim 1/3$ by approaching $\epsilon=1/3$, which
manifests the quite unique nature of the stripe order in the Sr-doped
nickelate system.\cite{katsufuji99,kaji01}

\begin{figure}
 \includegraphics[width=0.9\hsize]{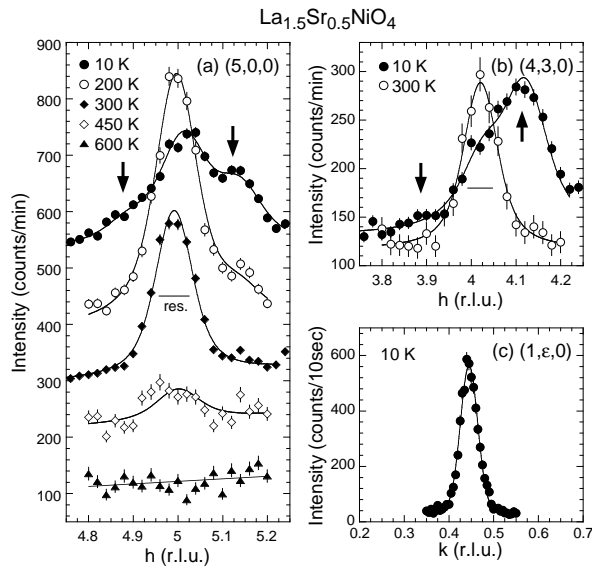}
 \caption{(a), (b) Temperature dependences of the profiles of charge
 order peaks at and around $(5,0,0)$ (panel (a)) and $(4,3,0)$ (panel
 (b)).  (c) Profile of a spin order peak $(1,0.44,0)$ at 10 K. In the
 panel (a), the $y$ axis is sifted by 100 counts for each
 data. Horizontal bars and solid lines in the panel (a) and (b) indicate
 the instrumental resolutions (FWHM) and Lorentzian profiles convoluted
 with the instrumental resolution, respectively. A solid line in the
 panel (c) is a fit to a Gaussian.}
 \label{fig_hscan}
\end{figure}

On the other hand, another well-known ground state of a hole doped
system is a so-called checkerboard-type charge order. The doping
concentration of 50\% holes is ideal for the checkerboard charge order,
and this charge order is widely seen in isostructural transition metal
oxides with $n_\mathrm{h}=1/2$ such as manganites \cite{sternlieb96} and
cobalates. \cite{zaliznyak00} Note that the $\epsilon=1/2$-limit of the
stripe order in nickelates coincides with the checkerboard charge
order. Early electron diffraction and transport
studies,\cite{chen93,cheong94} indeed, indicated the existence of a
checkerboard charge order at $n_\mathrm{h} = 1/2$.  From our recent
studies on the highly Sr-doped nickelate system, however, we failed to
detect any superlattice peaks from the incommensurate or checkerboard
charge order in the $x=1/2$ sample.\cite{yoshizawa00} Instead, we
observed an \textit{incommensurate spin} order with an
incommensurability $\epsilon \sim 0.44$ (see Fig.\
\ref{fig_hscan}(c)).\cite{yoshizawa00} Surprisingly, the stripe order
persists even at $n_\mathrm{h} =1/2$, and this result gives rise to a
question whether the ground state of the nickelates at $n_\mathrm{h}
=1/2$ is the checkerboard charge order or stripe charge order.

In the present study, we solve such a mystery of the charge ordering in
the 50\% doped nickelate, and demonstrate that the charge ordering shows
an interesting rearrangement from a checkerboard charge order to a
stripe order in the $x=1/2$ Sr-doped nickelate.  We found that the
checkerboard charge order is formed at $T \sim 480$ K, but a part of the
system changes over to the stripe-type charge order with decreasing $T$.
This complicated behavior is a consequence of a strong correlation of
the couplings between charge and spin degrees of freedom.

%\section{Experimental Procedures}

Large single crystal samples were grown by the floating zone
method. They were pre-characterized by resistivity measurements, and cut
into the size of $6\,\mbox{mm}\phi \times 25\,\mbox{mm}$ for the present
neutron diffraction study.  The neutron diffraction experiments were
performed using triple axis spectrometer GPTAS installed at the JRR-3M
reactor in JAERI, Tokai, Japan with a fixed incident neutron momentum of
$k_i =3.81\,\mbox{\AA}^{-1}$.  We chose a combination of horizontal
collimators of 40$^{\prime}$-40$^{\prime}$-40$^{\prime}$-80$^{\prime}$
(from monochromator to detector), and set two PG filters before the
inpile and after the sample positions to eliminate higher order
contaminations.  The temperature of the sample was controlled by a high
temperature-type closed-cycle He gas refrigerator within an accuracy of
0.2 K. Although the crystal has a tetragonal structure (space group
$I4/mmm$), we employ a larger unit cell of the size
$\sqrt{2}a_\mathrm{t} \times \sqrt{2}a_\mathrm{t} \times c_\mathrm{t}$
relative to the simple $I4/mmm$ lattice.

%\section{Results}

%\subsection{Fig. \ref{fig_resist}: Temperature dependence of the
%resistivity}

\begin{figure}
 \includegraphics[width=0.65\hsize]{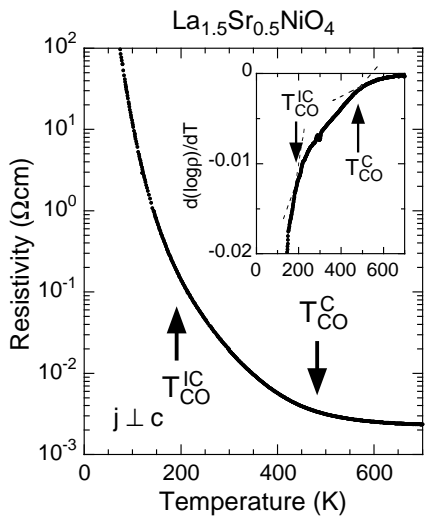}
 \caption{Temperature dependence of the resistivity within NiO$_{2}$
 plane (main panel) and its temperature derivative (inset).}
 \label{fig_resist}
\end{figure}

We first show the temperature ($T$) dependence of the in-plane
resistivity within the $ab$ plane ($\bm{j} \perp c$) of our
La$_{1.5}$Sr$_{0.5}$NiO$_{4}$ sample in Fig.\ \ref{fig_resist}. Though
the resistivity is almost metallic at high temperature
($\rho(600\,\textrm{K}) \sim 2.5 \times 10^{-3}$ $\Omega$cm), it still
increases monotonically as $T$ is lowered. A careful examination of the
$T$ derivative of the resistivity, $d\log \rho/dT$ (inset) suggests that
the $T$ dependence of the resistivity changes its slope twice: the first
clearer change locates at $T_\mathrm{CO}^\mathrm{C} \sim 480$ K, and the
second crossover-like change occurs at $T_\mathrm{CO}^\mathrm{IC} \sim
180$ K.  We note that the anomaly at $T_\mathrm{CO}^\mathrm{C} \sim 480$
K is substantially higher than the anomaly reported in early studies of
the resistivity and electron diffraction patterns,\cite{chen93,cheong94}
and that the one at $T_\mathrm{CO}^\mathrm{IC} \sim180$ K is recognized
for the first time in the present study.  It is possible that these
anomalies in the resistivity corresponds to a two-step process of the
charge localization, and they may indicate that two different types of
charge ordering are formed below the temperatures of two respective
anomalies.

%\subsection{Fig. \ref{fig_hscan}: In-plane profiles of the charge
% order peaks}

In order to find signals due to possible charge orderings inferred by
the anomalies of the resistivity, we surveyed reciprocal lattice
positions around $(h,k,0)$ with $h+k=\mbox{odd}$, where superlattice
peaks due to the checkerboard charge order are expected. Figures
\ref{fig_hscan}(a) and (b) show scan profiles along the [100] direction
around $(5,0,0)$ and $(4,3,0)$ at several $T$'s.  In Fig.\
\ref{fig_hscan} (c) is also shown a profile of a magnetic peak at
$(1,\epsilon,0)$ measured along the [010] direction at 10 K. In Figs.\
\ref{fig_hscan}(a) and (b), quite sharp peaks appear at commensurate
positions at $(5,0,0)$ and $(4,3,0)$ below
$T_\mathrm{CO}^\mathrm{C}$. These profiles clearly demonstrate that the
checkerboard charge ordering is formed in our $x=1/2$ sample below
$T_\mathrm{CO}^\mathrm{C}$, consistent with the early electron
diffraction experiment.\cite{chen93}

More importantly, a pair of satellite peaks emerges below
$T_\mathrm{CO}^\mathrm{IC}$ on both sides of the commensurate
checkerboard charge order superlattice peaks with further lowering $T$,
as indicated by arrows in Figs.\ \ref{fig_hscan}(a) and
\ref{fig_hscan}(b) ($h \sim 5 \pm 0.13$ and $h \sim 4 \pm 0.13$,
respectively).  With use of the incommensurability of the magnetic
reflections $\epsilon \sim 0.44$, the satellite peaks are indexed as $(h
\pm 2\epsilon, k,0)$ with $h + k = \mbox{even}$. Thereby, we identify
them as the charge and spin superlattice peaks from the stripe-type
charge order with $\epsilon \sim 0.44$. These observations lead us to
conclude that the two types of charge order coexist at low temperatures
in the Sr-doped nickelate with $n_\mathrm{h} =1/2$.

To determine correlation lengths within the NiO$_{2}$ planes, the
profiles of the charge order peaks in Fig.\ \ref{fig_hscan} are fitted
to a Lorentzian form with the instrumental resolution. The correlation
length for the checkerboard charge order becomes
$\xi_\mathrm{CO}^{\mathrm{C},h} \approx 40$ {\AA} at 300 K, but is
reduced to $\approx 20$ {\AA} at 10 K following the development of the
stripe charge order below $T_\mathrm{CO}^{\mathrm{IC}}$.  At 10 K, the
correlation length of the stripe charge order
$\xi_\mathrm{CO}^{\mathrm{IC},h}$ is $\approx 20$ {\AA}. Both
$\xi_\mathrm{CO}^{\mathrm{C},h}$ and $\xi_\mathrm{CO}^{\mathrm{IC},h}$
are nearly equal, but much shorter than the correlation length of the
spin order $\xi_\mathrm{SO}^{h} \approx 120$ {\AA} (10
K).\cite{yoshizawa00} It should be noted that
$\xi_\mathrm{CO}^{\mathrm{C},h}$ is also very close to the literature
value for the checkerboard charge order in
La$_{1.5}$Sr$_{0.5}$CoO$_{4}$, 26(2) {\AA}.\cite{zaliznyak00} We also
measured the profiles of the charge order peaks along the [001]
direction to examine the stacking of the charge order (not shown). We
found the scattering intensity of either type of charge ordering has a
very weak $l$ dependence. The profile of the checkerboard order exhibits
peaks at $l=\mathrm{even}$ positions, while that of the stripe order
shows a weak bump at $l=\mathrm{odd}$. From the observed profiles, we
estimate the out-of-plane correlation length of the checkerboard charge
order $\xi_\mathrm{CO}^{\mathrm{C},l}$ is $\sim 1$ {\AA} at 300 K and
$\sim 2$ {\AA} at 10K, while that of the stripe order
$\xi_\mathrm{CO}^{\mathrm{IC},l}$ is $\sim 3$ {\AA} at 10 K. For the
checkerboard charge order, the stacking along the [100] direction and
that along the [010] direction are equivalent, because the arrangement
of the charges is isotropic within NiO$_{2}$ planes. Quite short
correlation length of the spin and charge stripe order between NiO$_{2}$
planes is likely caused by a frustration of the stacking of the charge
stripes.

%\subsection{Fig. \ref{fig_Tdep}: Temperature dependencies of the charge
% order peaks and spin order peak.}

\begin{figure}
 \includegraphics[width=0.75\hsize]{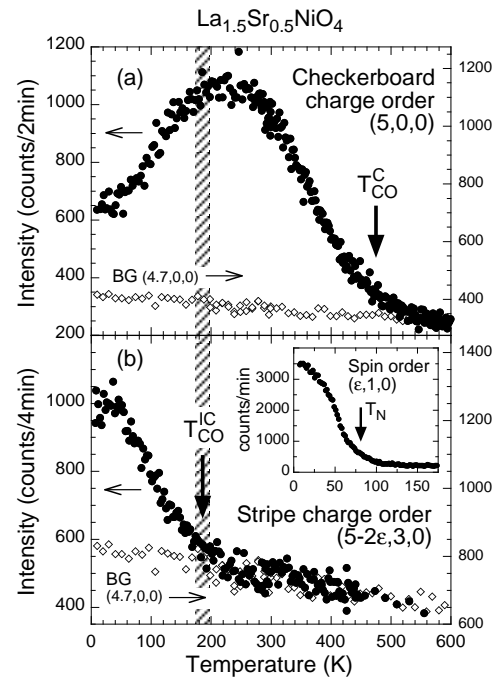}
 \caption{Temperature dependencies of the peak intensities of the
 commensurate charge order peak $(5,0,0)$ (a) and stripe charge order
 peak $(4.13,3,0)$ (b). The background intensities measured at
 $(4.7,0,0)$ are also depicted. The inset in (b) is the temperature
 dependence of the peak intensity of the stripe spin order peak at
 $(0.44,1,0)$.}
 \label{fig_Tdep}
\end{figure}

In order to examine an ordering process of the two charge orders, the
$T$ dependencies of the intensity of the commensurate peak at $(5,0,0)$
and the satellite peak at $(4.13,3,0) = (5-2\epsilon,3,0)$ are depicted
in Fig.\ \ref{fig_Tdep}. With decreasing $T$, the intensity at $(5,0,0)$
starts to grow around $T_\mathrm{CO}^\mathrm{C} \sim 480$ K.\cite{com1}
This temperature is in good agreement with the anomaly in the
resistivity (Fig.\ \ref{fig_resist}), but is much higher than the charge
ordering temperature reported in Ref.\ \onlinecite{chen93}. Further
decreasing $T$, the intensity reaches a maximum around
$T_\mathrm{CO}^\mathrm{IC} \sim 180$ K, and then turns to decrease below
$T_\mathrm{CO}^\mathrm{IC}$. The maximum of the commensurate component
is well correlated with the onset of the satellite (incommensurate)
components. The intensity of the satellite peak at $(4.13,3,0)$
gradually develops below $T_\mathrm{CO}^\mathrm{IC}$. Clearly, this
temperature coincides with the second anomaly in the resistivity shown
in Fig.\ \ref{fig_resist}.  In addition, the stripe spin order sets in
at much lower temperature $T_\mathrm{N} \sim 80$ K (see the inset of
Fig.\ \ref{fig_Tdep}(b)).  These $T$-dependent behaviors of the stripe
charge and spin order manifest the marvelous stability of the stripe
order at the commensurate hole concentration $n_\mathrm{h} \sim1/2$ in
the Sr-doped nickelate.

%\subsection{Fig. \ref{fig_diagram}: Phase diagram}

The robustness of the stripe order observed in the $x=1/2$ sample
motivated us to investigate the higher doping region for $x > 1/2$. From
our preliminary neutron diffraction measurements on the
Nd$_{2-x}$Sr$_{x}$NiO$_{4}$ system, we found that the well-defined
stripe-type order persisted even in the samples with the Sr
concentration as high as $x = 0.7$.  We also observed a weak signal of
the checkerboard charge ordering in the $x=0.6$ sample.  The observation
of the stripe ordering for $x \gtrsim 1/2$ is itself very unusual, and
the nature of this stripe ordering deserves further experimental and
theoretical studies.

\begin{figure}
 \includegraphics[width=0.75\hsize]{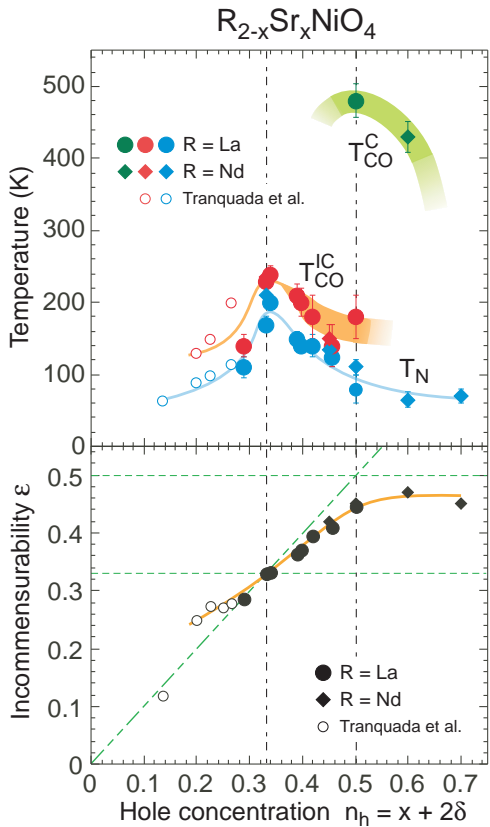}
 \caption{(color) Phase diagram of $R_{2-x}$Sr$_{x}$NiO$_{4}$: (a) Hole
 concentration dependence of the transition temperatures,
 $T_\mathrm{CO}^\mathrm{IC}$, $T_\mathrm{CO}^\mathrm{C}$ and
 $T_\mathrm{N}$. For details, see text. (b) Hole concentration
 dependence of the incommensurability $\epsilon$. Circles denote the
 data for La$_{2-x}$Sr$_{x}$NiO$_{4}$, while diamonds for
 Nd$_{2-x}$Sr$_{x}$NiO$_{4}$.  Open symbols are taken from Ref.\
 \onlinecite{tranquada96}.}
 \label{fig_diagram}
\end{figure}

In Fig.\ \ref{fig_diagram}, we summarize the ordering temperatures
$T_\mathrm{CO}^\mathrm{IC}$, $T_\mathrm{CO}^\mathrm{C}$ and
$T_\mathrm{N}$, and the incommensurability $\epsilon$ of the stripe
ordering in La$_{2-x}$Sr$_{x}$NiO$_{4+\delta}$ as well as
Nd$_{2-x}$Sr$_{x}$NiO$_{4}$ as a function of the hole concentration
$n_\mathrm{h}$. All of these values for the Nd samples are in excellent
accord with those for the La samples.  As reported
previously,\cite{yoshizawa00} $\epsilon$ shows a systematic deviation
from the linear relation of $\epsilon = n_\mathrm{h}$ around
$n_\mathrm{h} = 1/3$. Interestingly, we found that $\epsilon$ saturates
for $x \ge 1/2$ with the value $\epsilon \sim 0.44$ as shown in the
lower panel of Fig.\ \ref{fig_diagram}.  This result manifests that the
stripe state is robust against the hole doping even for $n_\mathrm{h}
\gtrsim 1/2$, and may indicate that the electronic state for $x > 1/2$
is similar to that of $x=1/2$.  The details of the charge and spin
ordering at $x > 1/2$ will be reported elsewhere.\cite{ishizaka_un}

%\section{Discussions}

%\subsection{Origin of the stripe phase}

\begin{figure}
 \includegraphics[width=\hsize]{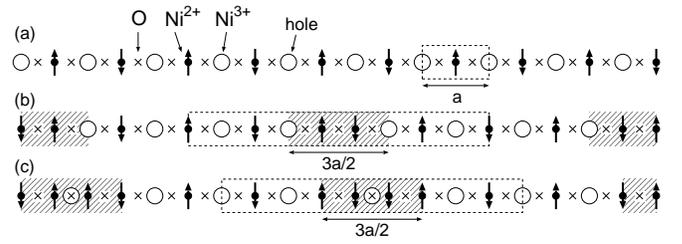}
 \caption{Model patterns of the charge ordering in the NiO$_{2}$ plane
 along the [100] direction for the commensurate charge ordered state (a)
 and the incommensurate charge ordered state for $\epsilon = 4/9$ (b), \
 (c). Small filled circles with arrows, open circles and crosses denote
 Ni$^{2+}$ sites, holes, and oxygen sites, respectively. Arrows indicate
 the spin directions of the Ni$^{2+}$ sites. Broken lines outline the
 unit cell of the charge ordering. The holes in the discommensuration
 region (shaded areas) occupy the metal sites in (b) while the oxygen
 sites in (c).}
 \label{fig_model}
\end{figure}

Next, we discuss the ordering process of the charge and spin ordering in
$x=1/2$.  The incommensurate charge ordering in $x=1/2$ can be
understood in terms of a discommensuration picture of the stripe
ordering introduced in our previous study.\cite{yoshizawa00} An example
of the stripe order near $n_\mathrm{h}=1/2$ is illustrated in Fig.\
\ref{fig_model}, based on the discommensuration picture.  The
checkerboard charge order consists of a matrix of the $\uparrow \times
\bigcirc \times \downarrow$ pattern where arrows represent the spin
direction on Ni$^{2+}$ sites, while $\times$ and $\bigcirc$ symbols
denote the oxygen sites and a hole on the Ni sites, respectively. The
width of the unit cell of this ordering pattern becomes $a$ as
illustrated in Fig.\ \ref{fig_model}(a).  In the next step, a
discommensuration will be introduced to such a pattern. For this
purpose, an object of size $3a/2$ containing one hole (shaded areas in
Figs.\ \ref{fig_model}(b) and \ref{fig_model}(c)) will be embedded for
every three checkerboard units.  The hole at each discommensuration
region may occupy either a Ni site as $\bigcirc \times \uparrow \times
\downarrow \times \bigcirc$ (Fig.\ \ref{fig_model}(b)) or an oxygen site
as $\uparrow \times \downarrow \otimes \downarrow \times \uparrow$
(Fig.\ \ref{fig_model}(c)). It should be noted that one cannot determine
whether a hole occupy a Ni site or an oxygen site in the
discommensuration region only from the present study, although
O-centered discommensuration may be preferred by the Coulomb repulsion
between the holes.  In any event, the introduction of the
discommensuration leads to an incommensurate (stripe) charge ordering
with the incommensurability $\epsilon = (3+1)/(2\times 3 + 3 \times 1) =
4/9 = 0.44$, being consistent with our observations.

%The checkerboard charge order consists of a matrix of the $\uparrow
%\times \bigcirc \times \downarrow$ pattern with the $a$-wide unit where
%arrows, $\times$, and $\bigcirc$ represent the spin direction on
%Ni$^{2+}$ sites, oxygen sites, and a hole on a Ni site, respectively.
%Within such a pattern, an object with a unit-width of $3a/2$ containing
%one hole (shaded areas in Figs.\ \ref{fig_model}(b) and
%\ref{fig_model}(c)) is embedded for every three checkerboard units,
%\cite{com2} and make an incommensurate charge ordering with $\epsilon =
%(3+1)/(2\times 3 + 3 \times 1) = 4/9 = 0.44$.

An important result of the discommensuration is the energy gain by the
spin exchange interactions. For the checkerboard charge ordered state,
the exchange interaction between the nearest neighbor sites is unique,
and there is only the one between Ni$^{2+}$ and Ni$^{3+}$ ions.  By
contrast, when the charge order becomes incommensurate, other kinds of
exchange interactions between Ni ions can be introduced in the
discommensuration regions: the Ni-centered discommensuration yields
Ni$^{2+}$$-$Ni$^{2+}$ bonds (Fig.\ \ref{fig_model}(b)) while
Ni$^{2+}$$-$$\mathrm{hole}$$-$Ni$^{2+}$ bonds are brought about by the
O-centered discommensuration (Fig.\ \ref{fig_model}(c)). Because the
exchange interactions for these bonds are expected to be much stronger
than that between Ni$^{2+}$ and Ni$^{3+}$,\cite{zaanen94} the energy
gain due to the spin exchange interactions is larger in the stripe
state, and favors the stripe order.

%An important result of the discommensuration is the energy gain by the
%spin exchange interactions. For the checkerboard charge ordered state,
%the exchange interaction between the nearest neighbor sites is unique,
%and there is only the one between Ni$^{2+}$ and Ni$^{3+}$ ions, which is
%expected to be quite small.\cite{zaanen94} By contrast, when the charge
%order becomes incommensurate, Ni$^{2+}$$-$Ni$^{2+}$ bond (Fig.\
%\ref{fig_model}(b)) and/or Ni$^{2+}$$-$$\mathrm{hole}$$-$Ni$^{2+}$ bond
%(Fig.\ \ref{fig_model}(c)) may be introduced in the discommensuration
%region, depending on whether the holes enter the metal site or the
%oxygen site. Because the exchange interaction for these bonds is much
%stronger than that between Ni$^{2+}$ and Ni$^{3+}$, the energy gain due
%to the spin exchange interactions is larger in the stripe state, and
%favors the stripe ordering.

From these considerations, a possible scenario of the ordering process
of the charge ordering in the 50\%-doped nickelate is as follows: At
high $T$ where the spin exchange interactions are disturbed by thermal
fluctuations, small polarons are created through the electron-phonon
coupling and they form the checkerboard order due to the Coulomb
interactions between polarons.\cite{zaanen94} With decreasing $T$,
however, the spin correlation develops, and eventually a large energy
gain of the Ni$^{2+}$$-$Ni$^{2+}$ bonds stabilizes the stripe order by
introducing holes into the checkerboard pattern. Although the
checkerboard charge order is favored by \textit{charges}, it is further
disfavored by the \textit{spin} exchange interactions. The latter favors
the incommensurate ordering and results in the coexistence of the
commensurate and stripe oredered regions at low temperatures in the
nickelate system for $x \gtrsim 1/2$. The present observation suggests
the importance of the role of the spin exchange interactions in the
charge ordering in the Sr-doped nickelate system.

It would be interesting to compare the present stripe formation at
$x=1/2$ with the appearance of the stripe phase in a cuprate with a very
low hole concentration. Very recently, Matsuda \textit{et al.}\ revealed
that a lightly-doped La$_{2-x}$Sr$_{x}$CuO$_{4}$ with $x < 0.02$ shows a
phase separation between a non-doped N\'{e}el ordered region and a
diagonal stripe-phase region with $\epsilon \sim 0.02$.\cite{matsuda02}
In the lightly-doped cuprates, the stripe with a modulation of $\epsilon
\sim 0.02$ emerges from the matrix of the non-doped N\'{e}el ordered
state, while in the nickelate with $x=1/2$, the stripes emerge from the
matrix of the checkerboard charge order, and result in an antiphase
domain modulation with a propagation vector of $1/2 -\epsilon \sim
0.05$.  In this sense, the present results for the $n_\mathrm{h} =1/2$
nickelate can be regarded as a counterpart of the behavior of the
cuprate with $n_\mathrm{h} \lesssim 0.02$.  We would like to suggest
that, in general, a stripe phase is magnetically favored more than the
state with a homogeneous hole distribution even in the limiting case of
0\% or 50\% doping.

%\section{Conclusion}

In conclusion, we observed in La$_{1.5}$Sr$_{0.5}$NiO$_{4}$ a
rearrangement of the checkerboard-type charge ordering to the
stripe-type one. With decreasing $T$, a checkerboard charge order is
formed below $T_\mathrm{CO}^\mathrm{C} \sim 480$ K. This charge order
is, however, taken over by the stripe-type charge order with an
incommensurability twice as large as that of the spin order below
$T_\mathrm{CO}^\mathrm{IC} \sim 180$ K. The stripe ordered phase
survives up to $x = 0.7$ in Nd$_{2-x}$Sr$_{x}$NiO$_{4}$. These results
evidence the unusual robustness of the stripe order, and suggest the
importance of the interactions between spins and charges in the Sr-doped
nickelate system.

\begin{acknowledgments}

The authors thank K. Chatani and K. Hirota for a loan of the high
temperature-type closed cycle He gas refrigerator.  This work was
supported by a Grant-In-Aid for Scientific Research from the Ministry of
Education, Culture, Sports, Science, and Technology, Japan and by the
New Energy and Industrial Technology Development Organization (NEDO) of
Japan.

\end{acknowledgments}


\begin{thebibliography}{99}
\bibitem{com1} Due to the short range nature of charge orderings, the
 scattering intensity of the superlattice peaks do not show clear
 anomalies through the transition temperatures. We define
 $T_\mathrm{CO}$'s as temperatures below which the intensity
 increases with logarithmic $T$ dependence (see Ref.\
 \onlinecite{kaji01}).
\end{thebibliography}
\end{document}